\documentclass[aps,prl,twocolumn,superscriptaddress,showpacs,10pt]{revtex4-1}
\usepackage{graphicx}
\usepackage{amssymb}
\usepackage{amsmath}
\pdfoutput=1

\begin{document}

\def\e{\varepsilon}
\def\f{\varphi}
\def\p{\partial}
\def\ba{\mathbf{a}}
\def\bA{\mathbf{A}}
\def\bb{\mathbf{b}}
\def\bB{\mathbf{B}}
\def\bD{\mathbf{D}}
\def\be{\mathbf{e}}
\def\bE{\mathbf{E}}
\def\bH{\mathbf{H}}
\def\bj{\mathbf{j}}
\def\bk{\mathbf{k}}
\def\bK{\mathbf{K}}
\def\bM{\mathbf{M}}
\def\bm{\mathbf{m}}
\def\bn{\mathbf{n}}
\def\bq{\mathbf{q}}
\def\bp{\mathbf{p}}
\def\bP{\mathbf{P}}
\def\br{\mathbf{r}}
\def\bR{\mathbf{R}}
\def\bS{\mathbf{S}}
\def\bu{\mathbf{u}}
\def\bv{\mathbf{v}}
\def\bV{\mathbf{V}}
\def\bw{\mathbf{w}}
\def\bx{\mathbf{x}}
\def\by{\mathbf{y}}
\def\bz{\mathbf{z}}
\def\Bn{\boldsymbol{\nabla}}
\def\Bo{\boldsymbol{\omega}}
\def\Br{\boldsymbol{\rho}}
\def\Bs{\boldsymbol{\hat{\sigma}}}
\def\bh{{\beta\hbar}}
\def\mA{\mathcal{A}}
\def\mB{\mathcal{B}}
\def\mD{\mathcal{D}}
\def\mF{\mathcal{F}}
\def\mG{\mathcal{G}}
\def\mH{\mathcal{H}}
\def\mI{\mathcal{I}}
\def\mL{\mathcal{L}}
\def\mO{\mathcal{O}}
\def\mP{\mathcal{P}}
\def\mT{\mathcal{T}}
\def\mU{\mathcal{U}}
\def\mZ{\mathcal{Z}}
\def\fr{\mathfrak{r}}
\def\ft{\mathfrak{t}}
\newcommand{\rf}[1]{(\ref{#1})}
\newcommand{\al}[1]{\begin{aligned}#1\end{aligned}}
\newcommand{\ar}[2]{\begin{array}{#1}#2\end{array}}
\newcommand{\eq}[1]{\begin{equation}#1\end{equation}}
\newcommand{\bra}[1]{\langle{#1}|}
\newcommand{\ket}[1]{|{#1}\rangle}
\newcommand{\av}[1]{\langle{#1}\rangle}
\newcommand{\AV}[1]{\left\langle{#1}\right\rangle}
\newcommand{\braket}[2]{\langle{#1}|{#2}\rangle}
\newcommand{\ff}[4]{\parbox{#1mm}{\begin{center}\begin{fmfgraph*}(#2,#3)#4\end{fmfgraph*}\end{center}}}

\newcommand{\degree}{\ensuremath{^\circ}}

\def\mr{m_{\perp}}
\def\ml{m_{\parallel}}
\def\hr{H_{\perp}}
\def\hl{H_{\parallel}}

\def\mb{(\mu+\alpha\nu)}
\def\nb{(\nu-\alpha\mu)}
\def\lb{(\lambda+\alpha\kappa)}
\def\kb{(\kappa-\alpha\lambda)}
\def\mn{\left|\bm\times\bz\right|}

\title{Magnetic bit stability: Competition between domain-wall and monodomain switching}

\author{Silas Hoffman}
\author{Yaroslav Tserkovnyak}
\affiliation{Department of Physics and Astronomy, University of California, Los Angeles, California 90095, USA}

\author{Pedram Khalili Amiri}
\author{Kang L. Wang}
\affiliation{Department of Electrical Engineering, University of California, Los Angeles, California 90095, USA }

\begin{abstract}
We numerically study the thermal stability properties of computer memory storage realized by
a magnetic ellipse. In the case of practical magnetic random-access memory devices, the bit can form a spin
texture during switching events. To study the energy barrier for thermally-induced switching, we develop
a variational procedure to force the bit to traverse a smooth path through configuration space between the points of stability. We identify textured configurations realizing domain-wall propagation, which may have an energy barrier less than that of the corresponding monodomain model. We contrast the emergence of such micromagnetic effects in thermal versus field-induced switching.
\end{abstract}

\maketitle

The mechanisms of tunnel magnetoresistance \cite{jullierePL75} and spin-transfer torque \cite{slonczewskiJMMM96,*bergerPRB96} offer a simple and localized method in which to read and write computer memory by applying electric currents rather than magnetic fields \cite{ikedaIEEEE07,*chenIEEEM10}.  The corresponding devices have the potential to improve write energy and speed and decrease bit area.  Equally as important is the lifetime of the bit; that is, the robustness of the bit against thermal fluctuations between the ``on" and ``off" states.  This is furnished by the energy barrier, $E_b$, separating the points of bimodal stability and typically characterized by the dimensionless thermal-stability parameter, $\Delta=E_b/k_B T$, wherein $k_B$ is Boltzmann constant and $T$ ambient temperature.

There are principally two classes of methods to experimentally extract this thermal barrier: stochastic and field induced.  In stochastic methods, the thermal barrier is extrapolated from the thermal fluctuations of the magnet between the bistable points.  Because these fluctuations are of the order of the bit lifetime at room temperature, switching is in practice assisted by a large temperature, a spin-transfer torque, or a magnetic field \cite{kochPRL04,*lacourAPL04,*yagamiAPL04,*higoAPL05}. However, at large temperatures, it is becoming difficult to extrapolate the system parameters, such as magnetic saturation and anisotropies, making this method unreliable.  Likewise, although the thermal stability can be extrapolated faithfully at low voltages upon the application of a spin-transfer torque, at high voltages the local temperature may be too high to rely on the equilibrium parameters of our system due to the injection of ``hot" electrons and the associated Ohmic losses. For deterministic field-induced switching, one applies an external magnetic field in the easy plane \cite{upadhyayaJAP11}. By determining the magnitude of the magnetic field at which the system is no longer bimodally stable, the energy barrier can be ascertained.  Following this procedure for all directions in the easy plane, one obtains a Stoner-Wohlfarth astroid curve \cite{stoner1948,*tannousEJP08}. In order to be viable, all these methods generally assume a single-domain model of the magnet.  In this model, there is only one planar path through configuration space between one point of stability and the other. However, when the characteristic length scale of the spin texture becomes comparable to the  major axis of the elliptical bit, a richer configuration space becomes energetically accessible, opening the possibility for a textured path with a lower energy barrier between the stable points. As a result, an extrapolation of the thermal barrier can be ambiguous.

\begin{figure}[pt]
\includegraphics[width=0.9\linewidth,clip=]{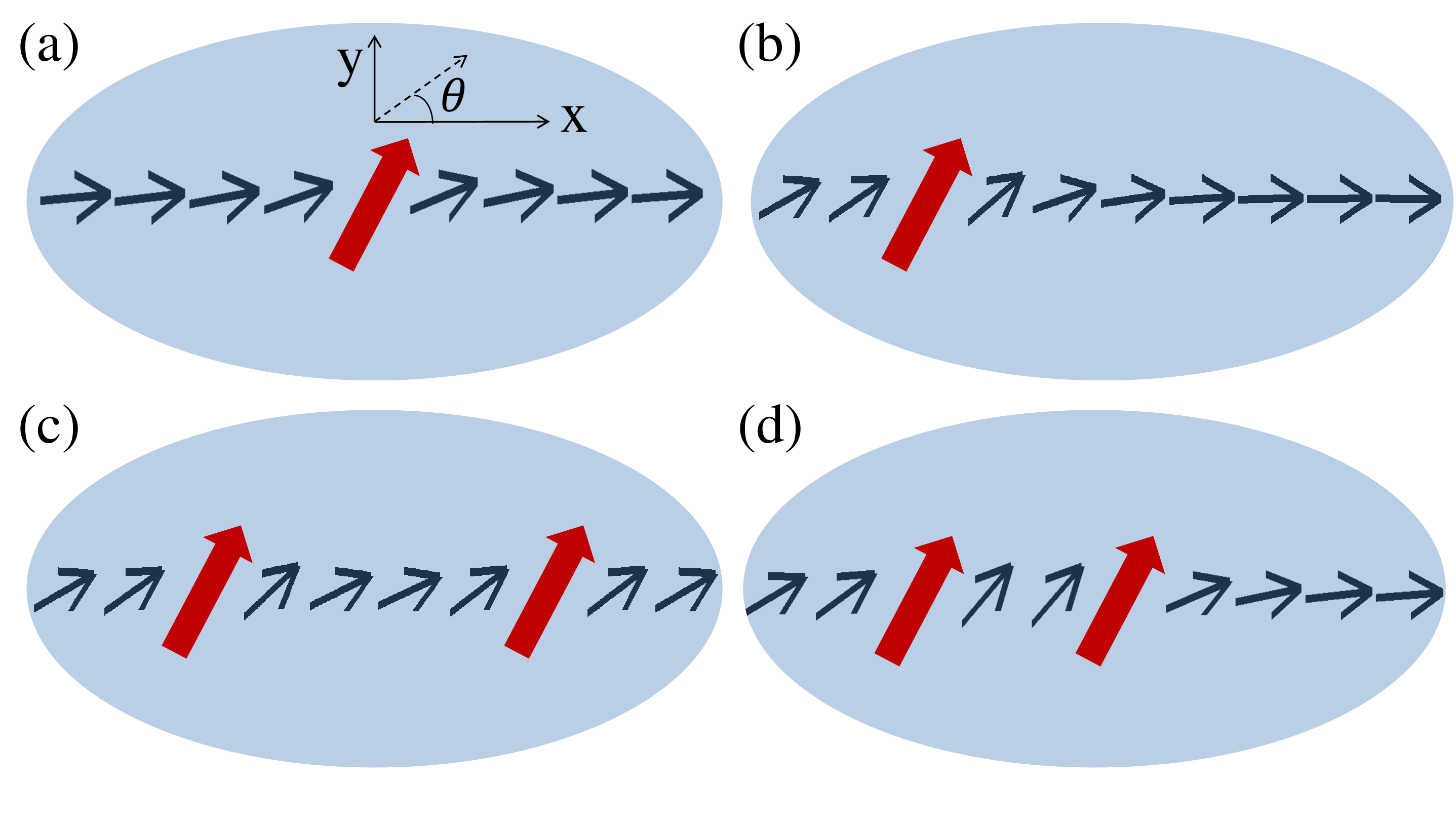}
\caption{Application of the local pinning fields (thick arrows) and schematic of the resulting magnetic textures (thin arrows): (a) Center pinning, (b) left off-center, (c) simultaneous left-off-center and right-off-center, and (d) simultaneous center and left-off-center. We find that (b) produces the lowest energy barrier.}
\label{fig1}
\end{figure}

In our analysis, we consider a magnetic ellipse of typical dimensions and material properties as those expected to be used as memory.  For these values, we find that the exchange energy is of the order of the demagnetization energy.  By applying slowly rotating local magnetic fields, which guide some variationally preselected points along the major axis (see Fig.~\ref{fig1}), we force the bit into a series of textured configurations through phase space that continuously evolve one stable minimum into the other.  We find a path between the bistable points, resembling domain-wall propagation, whose energy barrier is smaller than that of a single domain. For a concrete numerical demonstration,
we model our bit as a $V=150\times50\times2$~nm$^3$ thin-film ellipse with saturation magnetization of $M_s=1100$~emu/cm$^3$ and exchange stiffness $A=2$~$\mu$erg/cm. The crystalline anisotropy is neglected, and we consider only this single layer.  Our axes are oriented such that the $xy$ plane is in the plane of the ellipse and the easy (major) axis is along $x$. See Fig.~\ref{fig1}. The energy of the bit is written as sum of the exchange and demagnetization contributions:
\eq{F=E_{\rm xc}+E_{\rm dm}\,,}
where $E_{\rm xc}=A\int{dV (\nabla\bm)^2}$, $\bm$ is the unit magnetization vector, and $E_{\rm dm}$ is the dipole-dipole  interaction.  Because the applied magnetic field is used only to simulate a texture through phase space, in what follows we ignore the energetic contribution of the applied field. In the single-domain limit, $E_{\rm xc}=0$ and $E_{\rm dm}=(N_x m_x^2 + N_y m_y^2 + N_z m_z^2)VM_s^2$, where the demagnetization factors $N_x+N_y+N_z=2\pi$. The energy barrier is then simply given by $E_b=(N_y-N_x)V M_s^2$, corresponding to the $xy$-plane magnetization rotation. We define the characteristic length of the texture to be $\ell=\sqrt{A/(N_y-N_x)M_s^2}$.  Because for our structure $\ell\approx30$ nm, less than the major axis of our ellipse, micromagnetic (i.e., spin-texture) effects can be important. The latter are modeled using the \textit{LLG Micromagnetics Simulator} \cite{scheinfeinLLG}.

\begin{figure}[pt]
\includegraphics[width=\linewidth,clip=]{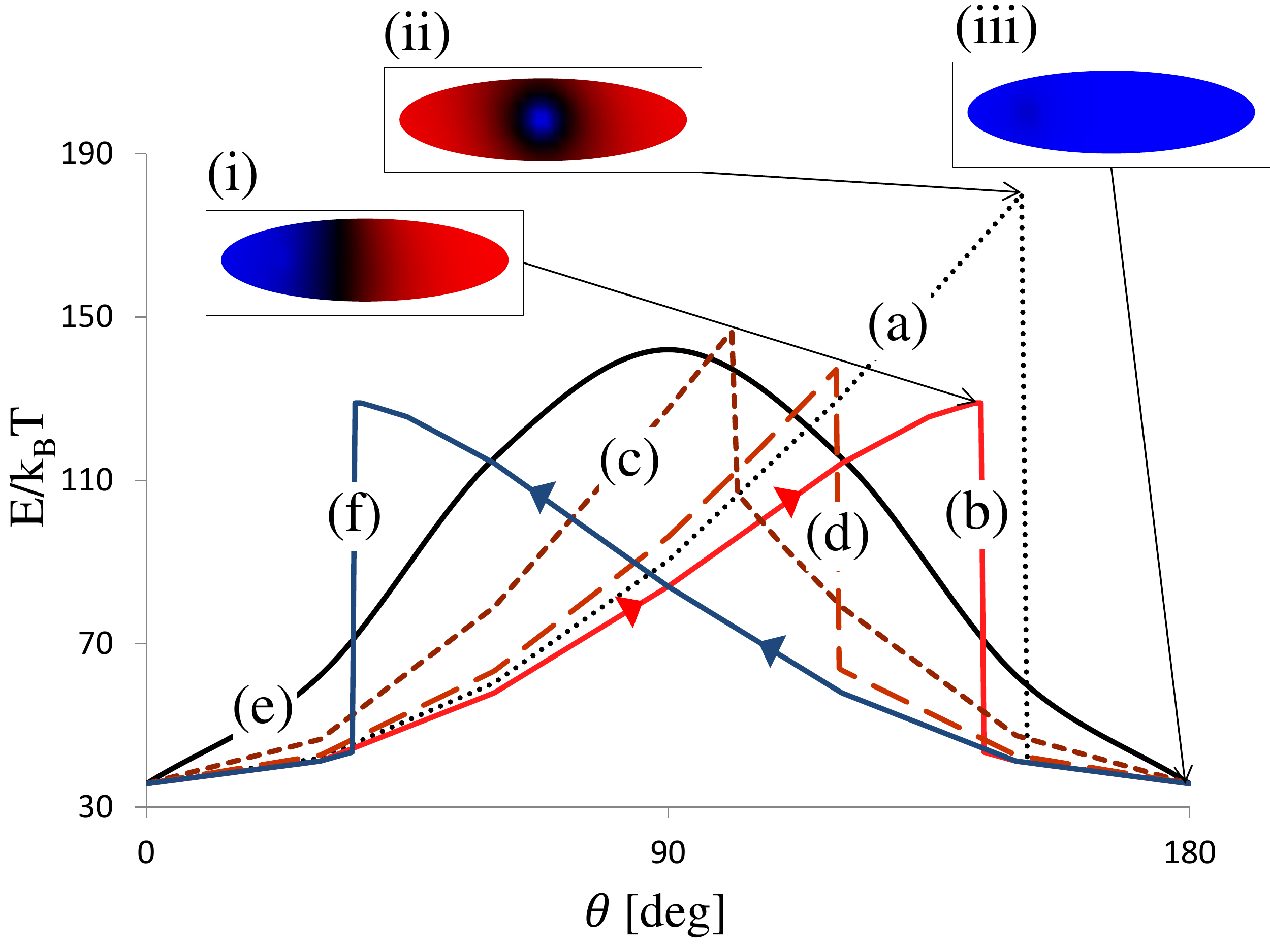}
\caption{Energy as a function of angle for different pinning prescriptions: (a), (b), (c), (d) of Fig.~\ref{fig1}, monodomain (e), and left off-center pinning while initializing the bit at $\theta=180\degree$ (f). Insets (i)-(iii) show $m_x$ at several critical points along the curves, with red corresponding to positive and blue negative magnetizations.}
\label{fig2}
\end{figure}

To find a continuous path through configuration space from one stable point to the other, we apply a single or a pair of localized pinning fields that force the magnet into a particular configuration, as sketched in Fig.~\ref{fig1}.  Initially oriented along the easy axis, we slowly rotate the pinning field in the $xy$ plane, allowing the (twisted) texture to equilibrate, and calculate the internal energy.  In order to variationally find an optimal path, we studied four different methods of applying the pinning field, which are depicted in Fig.~\ref{fig1}. Calculating the internal energy as a function of angle of the pinning fields, we are able to plot the energy of the system as the bit flips.  The principal results of our work can be found in Fig.~\ref{fig2}: Under application of a single off-center pinning, the energy barrier, $\Delta\approx90$, is less than that of a monodomain, $\Delta\approx110$. Since the thermal energy barrier is lowered by micromagnetic effects, the bit will follow a textured path in phase space when thermally flipping. Underlying the seemingly small change of $\lesssim20\%$ between energy barriers is thus a fundamental change in the mode of thermal switching. The Arrhenius thermal transition rate $\propto e^{-\Delta}$ is, furthermore, greatly reduced. Upon examining the texture dynamics, one finds that the left off-center pinning field nucleates a domain wall, while rotating the pinning field pushes the domain wall to the right along the $x$ axis.  This is most apparent at about $145\degree$, where, as illustrated in inset (i) of Fig.~\ref{fig2}, the two domains each occupy approximately half the area of the magnet.  At this point, the total internal energy is at a maximum, held in place only by the pinning field. After moving the domain wall past this point, there is a precipitous drop in internal energy as result of the domain-wall propagating to the right side of the magnet. This rapid drop in energy reflects a hysteretic process, wherein initially orienting the magnet at $0\degree$ or $180\degree$ with respect to the $x$ axis, the energy will respectively follow the (b) or (f) curve of Fig.~\ref{fig2}. We conclude that it is the domain-wall nucleation and propagation process that thermally flips the bit in our example. 

\begin{figure}[pt]
\includegraphics[width=0.9\linewidth,clip=]{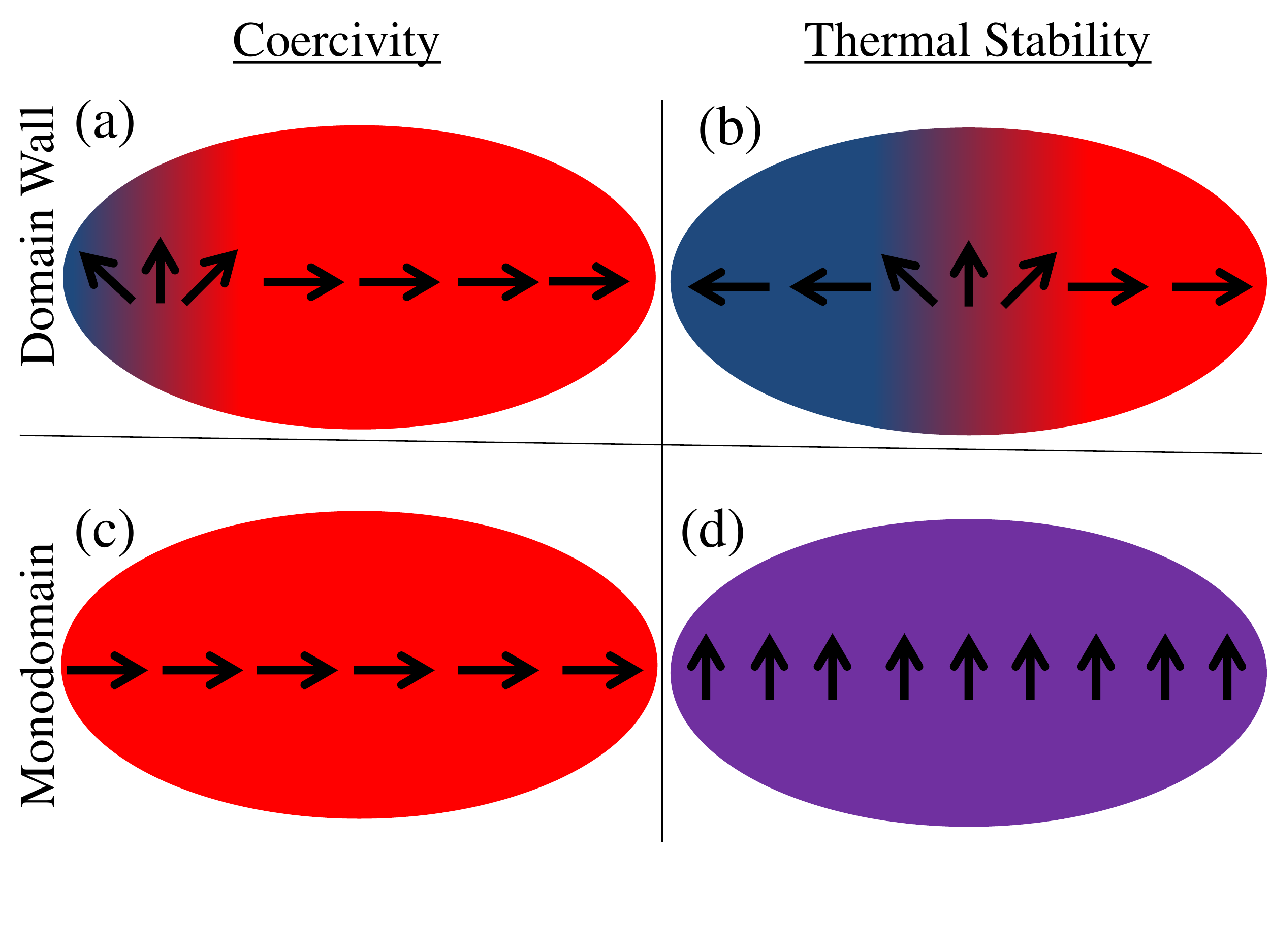}
\caption{Schematic illustration of the ``bottleneck" magnetic configurations for  switching under the application of magnetic field, (a) and (c), and during thermal fluctuations, (b) and (d).}
\label{fig3}
\end{figure}

To investigate the micromagnetic effects in the field-induced extraction of the thermal barrier, we apply a uniform magnetic field antiparallel to the direction of initialization of the bit.  Our magnetic bit flips at between 600 and 650~Oe, which is close to the value predicted by the single-domain anisotropy: 680~Oe. Closely examining the texture, we indeed see the magnet flips like a monodomain.  Analogous to the case of thin rectangular magnetic strips, wherein the coercivity field increases as the lateral width decreases, we suggest that the energetic bottleneck for domain-wall switching under application of an external field is at the tip of the bit where the domain wall nucleates. In contrast, the Arrhenius barrier for thermal domain-wall switching corresponds to the domain wall at the center of the magnet.  We can thus distinguish three principal scenarios subject to the characteristics of the bit: First, the bit flips like a monodomain under both thermal fluctuations and application of a coercivity field (as in the case of infinite stiffness), illustrated schematically in Figs.~\ref{fig3}(c) and (d), respectively.  Second, the bit flips thermally via domain-wall propagation [Fig.~\ref{fig3}(b)] and like a monodomain under application of a coercive field [Fig.~\ref{fig3}(d)] (as in our example). Third, the bit flips through domain-wall propagation both thermally and under application of a coercive field [Figs.~\ref{fig3}(a),(b)]. We find this in our simulations on more elongated ellipses, e.g., with dimensions $400\times80\times2$~nm$^3$ (and the same material parameters). The stochastic or field-induced characterization of a physical bit could correspondingly be significantly affected depending on the appropriate regime.

Since the exchange energy scales linearly with bit thickness while dipolar energy quadratically, upon increasing (decreasing) the thickness the texture effects are enhanced (reduced), and, therefore, the disparity in the thermal barrier between the single-domain path and the domain-wall should increase (decrease).  Particularly we take a thickness of 3~nm (1~nm) and find that the monodomain is $>20\%$ ($<1\%$) larger than the micromagnetic thermal barrier.

\begin{figure}[pt]
\includegraphics[width=\linewidth,clip=]{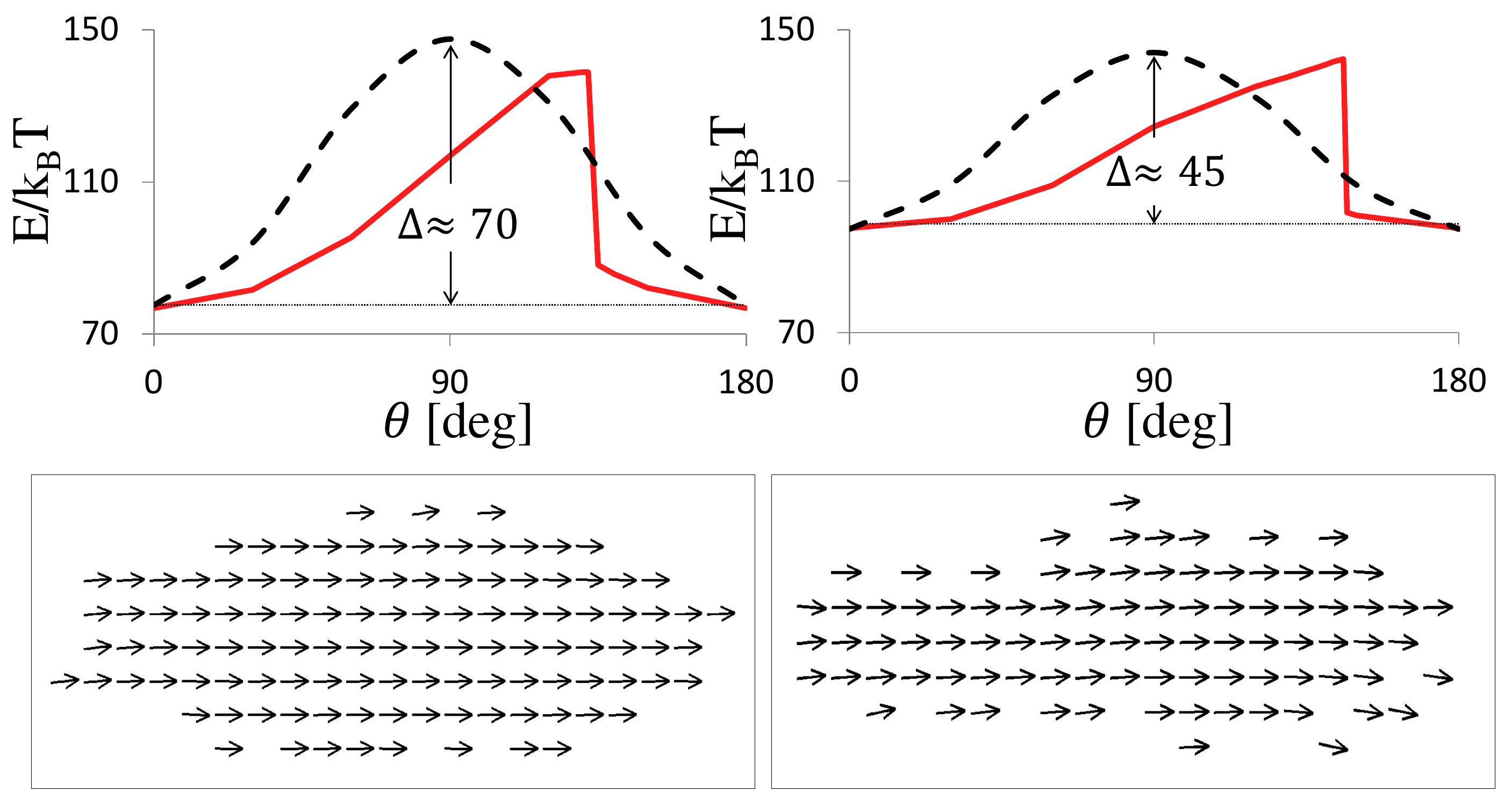}
\caption{Energy profile of bit with (a) edge roughness of amplitude 10~nm and wavelength 20~nm and (b) further enhanced roughness, where the dashed curve shows single-domain and the solid curve micromagnetic switching obtained under application of a local pinning field to the left off-center. The corresponding equilibrium bit textures are shown below.}
\label{fig4}
\end{figure}

We may expect random impurities and edge roughness to accentuate texture of the device and therefore affect the relative size of the thermal barrier in micromagnetic versus monodomain models.  As an example, we model roughness as a crude edge modulation with wavelength 20~nm and amplitude~10 nm, as depicted in the left half of Fig.~\ref{fig4}.  While this is perhaps larger than the characteristic edge roughness of a typical device, we take this as a somewhat exaggerated case.  Such edge modulation may, furthermore, be used to represent a generic structural or material disorder.  We find that the thermal energy associated with both modes (thermal and field-induced) of switching is diminished by disorder, while the relative difference between the single-domain and textured behavior, $\lesssim20\%$, is essentially unaffected [see Fig.~\ref{fig4}(a)]. Further distorting the edge (right half of Fig.~\ref{fig4}), the thermal barriers become nearly equal [see Fig.~\ref{fig4}(b)].  By way of roughening the edge, the aspect ratio of the elliptical bit has decreased, decreasing the $xy$ demagnetization factors and thus favoring monodomain switching. A systematic analysis of geometric and material imperfections falls outside  the scope of this work.

As a result of our variational optimization of the path taken through the configuration space, there may be a trajectory of states with even a lower associated thermal barrier. In other words, the domain-wall type of propagation discussed here produces the upper bound for the thermal barrier. Finding at least one micromagnetic path, however, whose thermal barrier is smaller than that of the monodomain, suggests a qualitatively different scenario of thermal flipping and a lower thermal barrier, compared to the monodomain. We find such effects, with regard to both thermal stability and astroid diagrams, ubiquitous in practical devices.

The authors would like to thank Juan G. Alzate, Scott Bender, Ilya Krivorotov, and Pramey Upadhyaya for their stimulating input. This work was supported in part by the DARPA, the NSF under Grant No. DMR- 0840965, and the Alfred P. Sloan Foundation (Y.T.).

\end{document}